\documentclass[preprint,showpacs,preprintnumbers,amsmath,amssymb,nofootinbib]{revtex4}
\usepackage{epsfig}

\begin{document}

\title{Continuum Coupling Effects in Spectra of Mirror Nuclei and Binding Systematics}

\author{J. Oko{\l}owicz$^{1,2}$}
\author{M. P{\l}oszajczak$^1$} 
\author{Yan-an Luo$^1$}

\affiliation{%
$^1$ Grand Acc\'{e}l\'{e}rateur National d'Ions Lourds (GANIL),
CEA/DSM -- CNRS/IN2P3, BP 5027, F-14076 Caen Cedex 05, France\\
$^2$ Institute of Nuclear Physics, Radzikowskiego 152,
PL - 31342 Krakow, Poland
}

\begin{abstract}
Continuum-coupling correction to binding energies in neutron rich oxygen and fluorine isotopes and to excitation energies of $2_1^+$ states in  $^{36}\mbox{Ca}$ and $^{36}\mbox{S}$ mirror nuclei are studied using the real-energy continuum shell model. 
 \end{abstract}
\maketitle

\section{Introduction}
Atomic nuclei form a network of coupled systems communicating with each other through decays and captures \cite{erice}. If continuum states are neglected, this communication is broken and each system becomes an isolated closed quantum system (CQS). It is obvious, that the CQS description of atomic nuclei (e.g. the nuclear shell model (SM)) becomes self-contradictory for weakly-bound or unbound states. 

A classic example of a continuum coupling is the Thomas-Ehrmann shift \cite{thomas} which manifests itself in the asymmetry in the energy spectra between mirror nuclei having different particle emission thresholds. A consistent description of the interplay between scattering states, resonances, and bound state requires an open quantum system (OQS) formulation. Comprehensive many-body theory of weakly bound/unbound states has been advanced recently in the time-asymmetric quantum mechanics using the complete ensemble of single-particle states  consisting of resonant (Gamow) states and the complex-energy, non-resonant continuum of scattering states from which the complete many-body basis of OQSs can be obtained \cite{gamow}. Another formulation of the continuum shell model is obtained by embedding standard SM in the continuum of decay channels. This approach provides a unified description of nuclear structure and nuclear reaction aspects \cite{karim,bnop2,luo,bnop3}. 

In this paper, we study the effect of the continuum coupling on spectra of $^{36}\mbox{Ca}$ and
 $^{36}\mbox{S}$ mirror nuclei. We show that the continuum coupling explains naturally not only the appearance of the asymmetry in spectra but also provides a major part of the Thomas-Ehrmann shift. We shall also discuss salient effects of a continuum coupling in the binding energy systematics, 
 in particular the anti-odd-even staggering (anti-OES) and the effective range of energies in which various decay channels are correlated with each other with the discrete many-body states changing significantly their energy and wave function.

\section{Shell model embedded in the continuum}
In the shell model embedded in the continuum (SMEC), nucleus is described as an OQS \cite{opr}. The total function space consists of the set of ${\cal L}^2$-functions, as in the standard SM, and the set of scattering states (decay channels). These two sets are obtained by solving the Schr\"odinger equation for discrete states  of the closed subsystem (closed quantum system (CQS)):
$H_{SM}\Phi_i=E_i^{(SM)}\Phi_i~,$
and for scattering states of the external environment:
$\sum_{c'}(E-H_{cc'})\xi_E^{c'(+)}=0~,$
where $H_{SM}$ is the SM Hamiltonian, and $H_{cc^{'}}=H_0+V_{cc^{'}}$ is the coupled-channel (CC) Hamiltonian. Channels: $c\equiv[J_i^{A-1};(l, j)]^{J_k^{A}}$,  are determined by the motion of an unbound particle with orbital angular momentum $l$ and total angular momentum $j$ relative to the residual nucleus with $A-1$ bound particles in a SM state $\Phi_j^{A-1}$. 
 $\xi_E^{c(+)}$ are channel projected scattering states with outgoing asymptotics. States of the daughter nucleus are assumed to be stable with respect to the particle emission. 
By means of two functions sets: ${\cal Q}\equiv \{\Phi_i^{A}\}$, ${\cal P}\equiv \{\zeta_E^{c(+)}\}$, one can define the corresponding projection operators
and the projected Hamiltonians: ${\hat Q}H{\hat Q}\equiv H_{QQ}$ and ${\hat P}H{\hat P}\equiv H_{PP}$ \cite{karim,opr}.  $H_{SM}$ is identified with the CQS Hamiltonian $H_{QQ}$  and $H_{cc}$ with 
$H_{PP}$. The coupling term $H_{PQ}$ is given by the two-body residual interaction \cite{opr}.

Schr\"odinger equation in the total function space splits into the two equations for projected operators $H_{QQ}$, $H_{PP}$. Assuming ${\cal Q}+{\cal P}=I_d$, one can determine a third wave function $\omega_i^{(+)}$ which is a continuation of SM wave function $\Phi_i^A$ in the scattering continuum. 
$\omega_i^{(+)}$ is obtained by solving the CC equations with the source term \cite{opr}.
Using the three function sets: $\{\Phi_i^A\}$, $\{\zeta_E^{c(+)}\}$ and $\{\omega_i^{(+)}\}$, one obtains a solution in the total function space:
\begin{eqnarray}
\label{sol1}
\Psi_E^c=\zeta_E^c+\sum_{i,k}(\Phi_i^A+\omega_i^{(+)}(E))\langle\Phi_i^A|
(E-{\cal H}_{QQ}(E))^{-1}|\Phi_k^A\rangle\langle\Phi_k^A|H_{QP}|\zeta_E^c\rangle
\end{eqnarray}
where $E$ is the total energy and ${\cal H}_{QQ}(E)$ is the 
energy-dependent effective Hamiltonian in ${\cal Q}$ subspace:
\begin{eqnarray}
\label{SMeq3}
{\cal H}_{QQ}(E)=H_{QQ}+H_{QP}G_P^{(+)}(E)H_{PQ}~ \ ,
\end{eqnarray}
where $G_P^{(+)}(E)$ in (\ref{SMeq3}) is the Green function in ${\cal P}$. ${\cal H}_{QQ}$ takes into account a modification of the CQS Hamiltonian ($H_{QQ}$) by couplings to the environment of decay channels.  ${\cal H}_{QQ}$ is a complex-symmetric matrix above the particle-emission threshold $E^{\rm (thr)}$ and Hermitian below it.  Diagonalization of ${\cal H}_{QQ}$ by an orthogonal and, in general, non-unitary transformation yields complex eigenvalues ${\tilde E}_i - \frac{1}{2}i{\tilde \Gamma}_i$, which depend on the energy $E$ of the particle in the continuum.  Energies and widths of the resonance states follow from:
$E_i={\tilde E}_i(E=E_i)$, $\Gamma_i={\tilde \Gamma}_i(E=E_i)$, where ${\tilde E}_i(E)$ and ${\tilde \Gamma}_i(E)$ are the eigenvalues of ${\cal H}_{QQ}(E)$. Details of the SMEC calculations can be found in Refs. \cite{karim,opr}.

\section{Features of the continuum-coupling energy correction to eigenvalues of the closed quantum system}
\label{bindsys}
In this chapter, we shall discuss salient features of the continuum-coupling energy correction:
 $E^{\rm (corr)}_i(E)=\langle\Phi_i|H_{QP}G_P^{(+)}(E)H_{PQ}|\Phi_i\rangle$, for the ground state (g.s.) of neutron-rich oxygen and fluorine isotopes. Details of the SMEC calculations for continuum-coupling energy correction to binding energies can be found in Ref. \cite{luo}. In these calculations, all possible couplings of the g.s. of $A$ nucleus to the states of $A-1$ nucleus are taken into account incoherently. 

\subsection{Binding systematics in neutron-rich nuclei}
For $T=1$ couplings, the average behavior of $E^{\rm (corr)}_{\rm gs}$ is determined by the strong dependence on $E_n^{\rm (thr)}$ (cf Fig. 1a of Ref. \cite{luo} for oxygen isotopes)\footnote{In the chosen $sdfp$ model space, $np$ continuum couplings are absent in oxygen isotopes.}. Close to the neutron drip line, this leads to an effective enhancement of $nn$ continuum-coupling strength which cannot be compensated by the $E_n^{\rm (thr)}$-independent correction of monopole terms. 
On the top of this behavior, one can see the odd-even staggering (OES) of $E^{\rm (corr)}_{\rm gs}(N)$. Blocking of the virtual scattering to continuum states by an unpaired nucleon diminishes the $nn$ continuum-coupling energy correction in odd-$N$ nuclei. For a fixed $E_n^{\rm (thr)}$, the coupling of 
$[N=2k,Z]$ system to  $[N=2k-1,Z]\otimes n$ decay channels is enhanced and the coupling of 
$[N=2k+1,Z]$ system to $[N=2k,Z]\otimes n$ decay channels becomes weaker with respect to an averaged behavior. This drip-line effect is seen only in a narrow range of excitation energies around $E_n^{\rm (thr)}(N)\simeq 0$ and  vanishes for $E_n^{\rm (thr)}(N)\geq 4$ MeV. For more realistic values of separation energies, as given by nuclear SM \cite{luo},  $E_n^{\rm (thr)}$ exhibits the pairing induced OES which becomes partially compensated close to a drip line by the anti-OES effect induced by couplings to decay channels, both opened and closed. 

In fluorine isotopes (cf Fig. 2a of Ref. \cite{luo}), $E^{\rm (corr)}_{\rm gs}$ is dominated by 
$np$ continuum couplings ($T=0,1$). The strength of $np$ continuum coupling
can be deduced by comparing SMEC results for  binding energies with experimental data \cite{luo}. 
An optimal value of $V_0^{(np)}/V_0^{(nn)}$ close to the neutron drip line
 is $V_0^{(np)}\simeq (1/2)V_0^{(nn)}$, whereas for nuclei close to the valley of beta-stability a standard choice is:  $V_0^{(np)}\simeq 2V_0^{(nn)}$.  Since the $np$ couplings provide a major part of this correction, therefore the gradual reduction of  $V_0^{(np)}/V_0^{(nn)}$ toward the neutron drip line 
leads to a non-linear dependence of the $E^{\rm (corr)}_{\rm gs}$ and, hence, to the dependence of two-body monopole terms on the neutron number. A similar dependence is expected if effective three-body interactions are included in the two-body framework of the SM \cite{zuker1}. 

The $np$ continuum coupling in odd-odd ($[N=2k+1,Z=2m+1]$) fluorine isotopes is increased as compared to the neighboring even-odd ($[N=2k+2,Z=2m+1]$ and $[N=2k,Z=2m+1]$) 
nuclei. Contrary to the continuum-coupling correction for like particles ($nn$ or $pp$ couplings), the characteristic anti-OES of $E^{\rm (corr)}_{\rm gs}$ persists even for $E_n^{\rm (thr)}>4$ MeV.
This effect is further enhanced by the OES of one-neutron (1n) emission thresholds which yields lower $E_n^{\rm (thr)}$ and, hence, larger continuum-coupling correction for odd-$N$ systems. 
The $np$ continuum-coupling energy correction attenuates the OES and can even wash it out close to drip lines if the ratio $V_0^{(np)}/V_0^{(nn)}$ would not be strongly reduced from its accepted value $\simeq 2$ in well-bound nuclei. 

In BCS formalism, the OES is associated with the blocking of a quasi-particle state close to the Fermi energy by an unpaired neutron ( resp. proton) what weakens $nn$ (resp. $pp$) pairing correlation in odd-$N$ (resp. odd-$Z$) isotopes. The proximity of continuum states in weakly-bound nuclei makes the blocking mechanism less effective, reducing the OES of one-nucleon separation energies. This reduction appears even though the strength of $nn$ ($pp$) pairing correlations is essentially unchanged. 

\subsection{Anatomy of the continuum-coupling correction}
\label{sec:anatomy}
\begin{figure}[h]
\centerline{\includegraphics[height=9.5cm,width=11cm,angle=00]{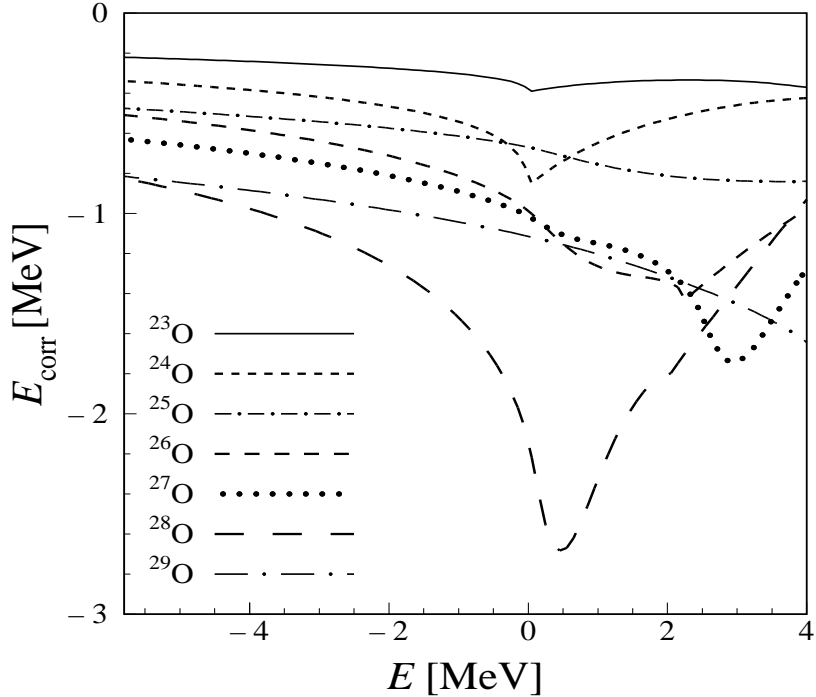}}
\caption{\small{The continuum-coupling energy correction to the SM g.s.
energy for different oxygen isotopes is plotted as a function of the neutron energy. Contributions from couplings to all available states in the daughter nucleus are added incoherently.}}
\label{fig:fign1}
\end{figure}

A typical behaviour of the total continuum-coupling correction $E^{\rm (corr)}_{\rm gs}$ to the g.s. SM 
energy (the CQS eigenvalue) for different oxygen isotopes is presented in Fig. \ref{fig:fign1} as a function of the neutron energy.  This correction is an incoherent sum of contributions from
couplings to all SM states in the daughter nucleus. $E=0$ corresponds to a position of the first  1n emission threshold. Rapid change of $E_{\rm corr}$ related to opening of next 1n emission threshold can be seen for $^{27}$O at $E\simeq 3$ MeV. The continuum-coupling energy correction rises with number of valence neutrons and in general is bigger in even-$N$ isotopes. Few notable exceptions can be seen right after the closure of the $sd$ shell (cf $^{29}$O in Fig. \ref{fig:fign1}) and at the begining of the $sd$ shell.
Behavior of the continuum-coupling correction depends on $l$ of the neutron wave involved  in a decay channel $[J_i^{\pi;A-1};(l,j)]^{J_{\rm g.s.}^{\pi;A}}$. This correction is largest exactly at the threshold of a channel $[J_i^{\pi;A-1};(l,j)]^{J_{\rm g.s.}^{\pi;A}}$ only for $l=0$ neutrons. For higher $l$-values or  for protons, the centrifugal barrier  and/or the Coulomb barrier shift the maximum of energy correction above the threshold. This can be seen for $^{28}$O, where $l=2$ neutron wave dominates in the g.s.\ to g.s.\ coupling. 

Couplings to excited states in $A-1$ nucleus are less important in even-$N$ isotopes. On the contrary,  in odd-$N$ isotopes they dominate. The distribution of contributions to $E^{\rm (corr)}_{\rm gs}$ as a function of the energy of corresponding states in $A-1$ nucleus, reflects the nature of pairing correlations in odd-$N$ and even-$N$ isotopes. Strong dissimilarity of g.s.\ to g.s.\ couplings between odd-$N$ and even-$N$ oxygen isotopes is attenuated if couplings to decay channels involving excited states in $A-1$ nucleus are taken into account. 
\begin{figure}[h]
\psfig{file=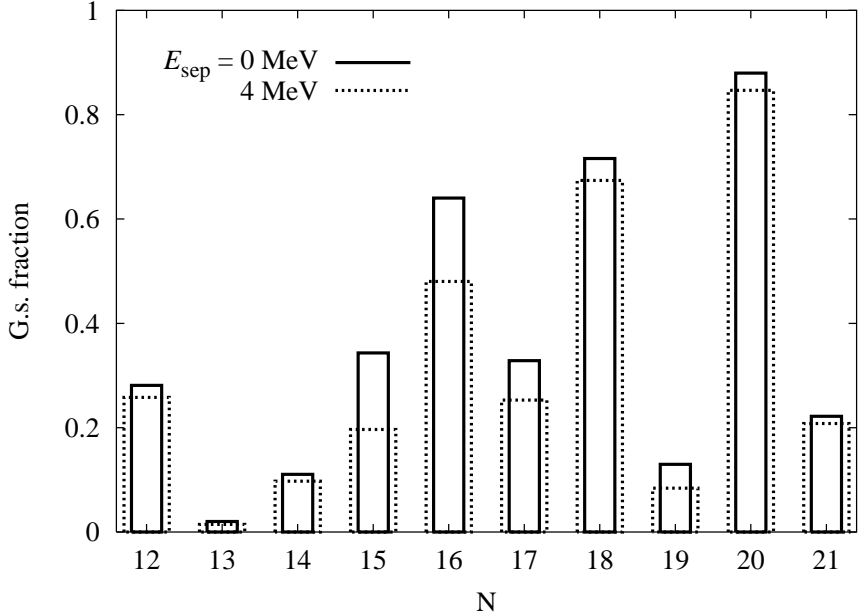,width=8.5cm,angle=-90}
\caption{\small{The fraction of the total continuum-coupling energy correction $E^{\rm (corr)}_{\rm gs}$ 
in oxygen isotopes, coming from a coupling to the g.s. of a daughter nucleus.  SMEC calculations are performed for fixed 1n-threshold energies: $E_n^{\rm (thr)}=0,4$ MeV.}}
\label{fig:fign3o}
\end{figure}
\begin{figure}[h]
\psfig{file=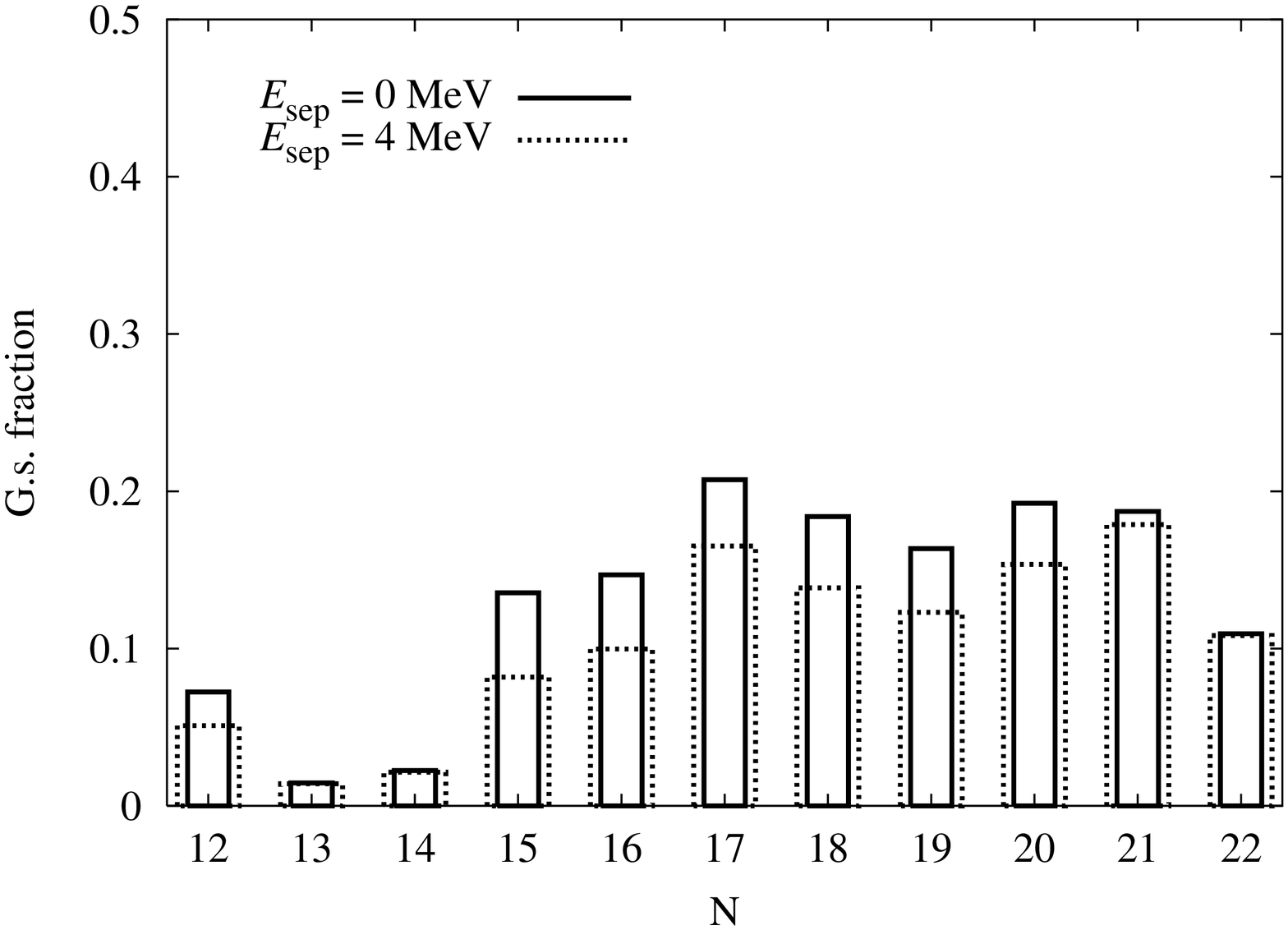,width=8.5cm,angle=-90}
\caption{\small{The same as in Fig. \ref{fig:fign3o} but for fluorine isotopes.}}
\label{fig:fign3}
\end{figure}
The fraction of $E^{\rm (corr)}_{\rm gs}$ coming from coupling 
to the g.s.\ of daughter nucleus is shown in Figs.~\ref{fig:fign3o} and \ref{fig:fign3} for oxygen and fluorine isotopes, respectively. In oxygen isotopes ($T=1$ couplings), one can see a strong OES following the staggering of neutron pairing correlations in these isotopes. In general, strong $nn$ pairing correlations in even-$N$ isotopes increase the weight of a g.s.\ to g.s.\ contribution. On top of this effect, one sees different  values of the g.s. fraction at the beginning of the shell (filling of $d_{5/2}$ subshell) and at the end of the shell (filling of $s_{1/2}$ and $d_{3/2}$ subshells). This picture changes qualitatively in fluorine isotopes (cf Fig. \ref{fig:fign3}). As pointed out in Sect. \ref{bindsys}, $np$ continuum coupling dominates in this isotopic chain, removing most of the OES due to the $nn$ correlations. As in oxygen isotopes, the fraction of $E^{\rm (corr)}_{\rm gs}$  coming from coupling to the g.s.\ of $A-1$ nucleus is smaller in $d_{5/2}$ subshell than in $s_{1/2}$ and $d_{3/2}$ subshells. Maximal value of this fraction is less than $\sim 20\%$ in fluorine isotopes, whereas for $^{24}$O, $^{26}$O and $^{28}$O
the g.s. fraction is $>60\%$.

\section{Mirror-symmetry breaking effect of a continuum coupling: Example of $2_1^+$ states in $^{36}$Ca and $^{36}$S}
Symmetry breaking effects in spectra of mirror nuclei are directly or indirectly related to the Coulomb interaction. Direct effects (the Thomas-Ehrmann effect) have been extensively discussed in the literature. Much less known indirect effects of the Coulomb interaction result from different positions of $n$/$p$ thresholds which modify continuum couplings effects and, hence, change spectra in mirror systems \cite{opr}. Recently, the energy of $2^+_1$ state has been measured in a weakly-bound $^{36}$Ca \cite{buerger}. An excitation energy of this proton-unbound state differs by $\sim 240$ keV from an energy of the well-bound mirror state in $^{36}$S. In this section, we will discuss in SMEC  the mirror-symmetry breaking continuum-coupling effects for these states. Details of SMEC calculations and the choice of an effective interaction are the same as described in Sect. \ref{bindsys} and in Ref. \cite{luo}.

SMEC excitation energies of $2_1^+$ states in $^{36}\mbox{Ca}$ and
$^{36}\mbox{S}$ are presented in the third column of Table \ref{table1}. The difference of excitation energies in those mirror  states is close to the experimental value (cf the first column of Table \ref{table1}), but absolute excitation energies are $\sim 0.5$~MeV bigger than experimental ones and $\sim 1$~MeV bigger than a SM value (cf the second column of Table \ref{table1}). One should notice that proton continuum couplings for $^{36}\mbox{Ca}$ and neutron continuum couplings for
$^{36}\mbox{S}$ result in a small difference of $2_1^+$ excitation energies in  $^{36}\mbox{Ca}$ and
$^{36}\mbox{S}$.  Almost a whole difference of $2_1^+$ excitation energies is due to the neutron-continuum couplings for  $^{36}\mbox{Ca}$ and proton-continuum couplings for $^{36}\mbox{S}$. This is due to the same structure of protons in $^{36}\mbox{Ca}$ and neutrons in $^{36}\mbox{S}$ for both $0_1^+$ g.s.\ and $2_1^+$ excited states (the shell closure). 
\begin{table}[h]
\caption{Excitation energy of $2_1^+$ states in $^{36}\mbox{Ca}$ and $^{36}\mbox{S}$ nuclei. 
Column labelled `SM' contains the SM results. The following three columns present SMEC results with different couplings. The third column shows results with both neutron and proton continua included. The next columns give the $2_1^+$ energy with only proton or neutron continuum couplings included.  All energies are in MeV.}
\label{table1}
\begin{center}
\begin{tabular}{|c|c|c|c|c|}
\hline
 Exp($^{36}$Ca)  & SM($^{36}$Ca) & SMEC($^{36}$Ca)  & $^{36}$Ca[$^{35}$K + p] & $^{36}$Ca[$^{35}$Ca + n] \\
 \hline 
& &   &   &   \\
3.05 $\pm$ 0.05 & 2.641 & 3.403 & 3.019 & 3.514  \\
&  &   &   &   \\
 \hline 
 \hline 
 Exp($^{36}$S)  & SM($^{36}$S) & SMEC($^{36}$S) & $^{36}$S[$^{35}$S + n] & $^{36}$S[$^{35}$P + p]  \\
  \hline 
& &   &   &   \\
3.2909 &  2.641 & 3.690 & 3.024 & 3.822 \\
 &   &   &   &   \\
  \hline  
 \end{tabular}
\vskip 0.3truecm
\end{center}
\end{table}  

\begin{figure}[h]
\begin{center}
\psfig{file=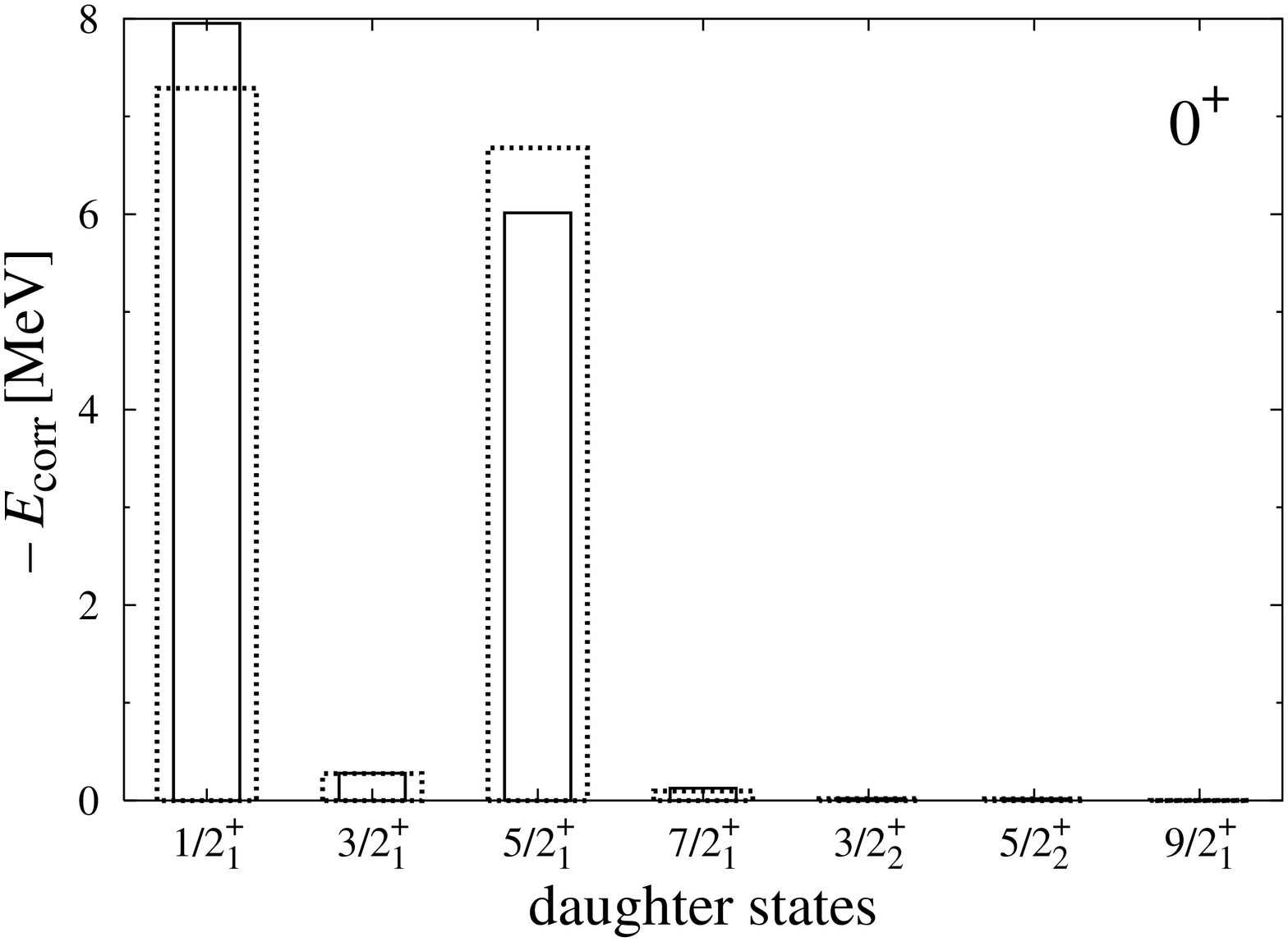,width=8cm,angle=-90}

\psfig{file=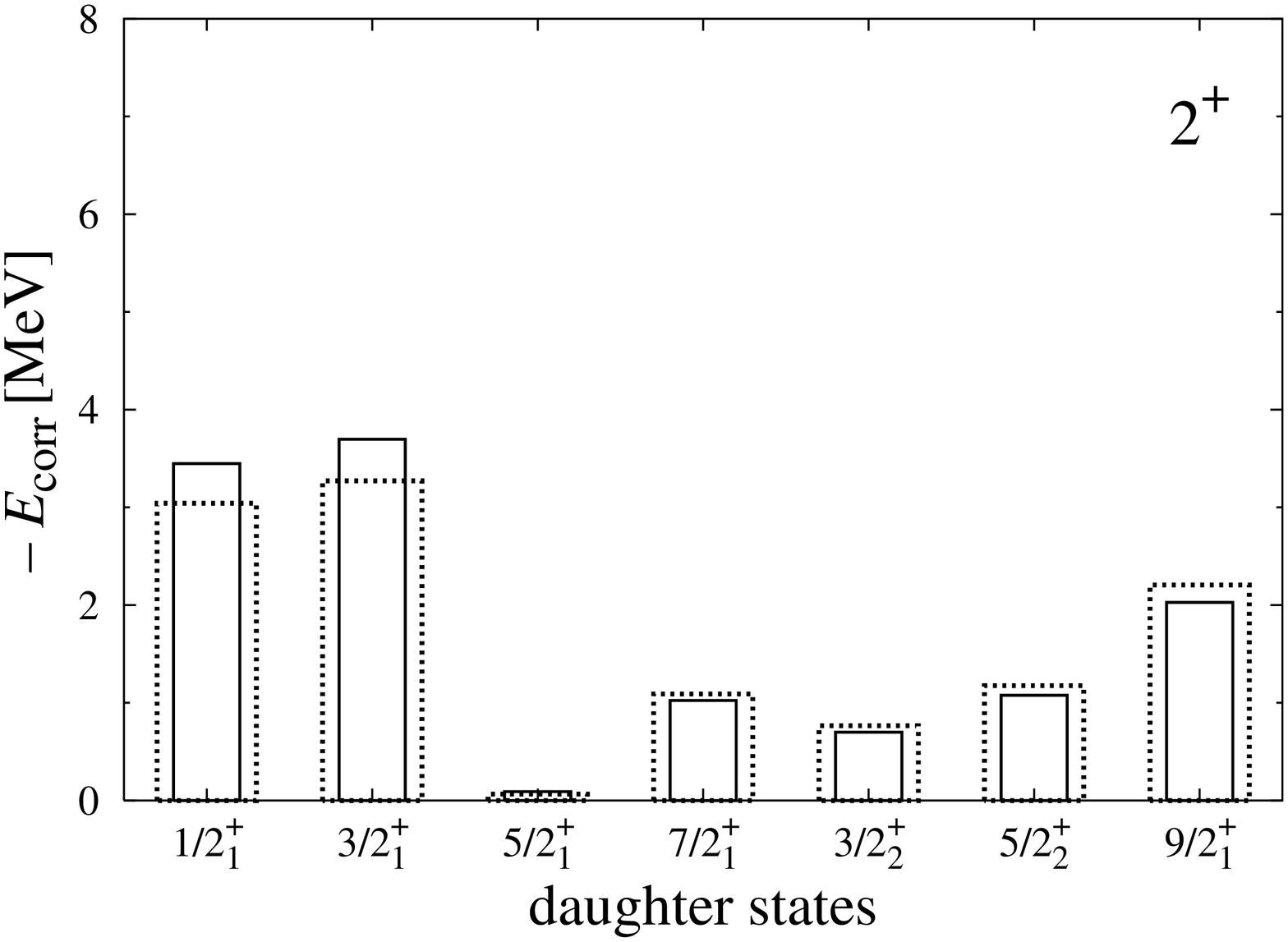,width=8cm,angle=-90}
\end{center}
\caption{The states $J_i^{\pi;A-1}$ in channel wave-functions $[J_i^{\pi;A-1};(l,j)]^{J^{\pi;A}}$ 
($J^{\pi;A}=0_1^+, 2_1^+$) which contribute most to the continuum-coupling correction from neutron continuum in $^{36}\mbox{Ca}$ 
($^{35}\mbox{Ca}$ states -- red narrow boxes) and proton continuum in $^{36}\mbox{S}$ 
($^{35}\mbox{P}$ states -- green wide boxes). The upper (lower) panel is for $0_1^+$ ($2_1^+$) state.}
\label{fig1}
\end{figure}

Let us take a closer look at the individual contributions from couplings via continuum to different states in $A-1$ nuclei. In Fig.~\ref{fig1}, contributions to the continuum-coupling energy corrections for $0_1^+$ and $2_1^+$ states in $^{36}\mbox{Ca}$ and $^{36}\mbox{S}$ are shown for selected daughter states. For $0_1^+$ state, the dominant energy contributions from couplings to the g.s. $1/2^+_{1}$ \ and the second excited state $5/2^+_{1}$ of $A-1$ nucleus are almost equal
\footnote{The balance of $E^{\rm (corr)}_{\rm gs}$ in $^{36}\mbox{Ca}$ and $^{36}\mbox{S}$ is a direct consequence of the separation energies in these nuclei: $S_p(^{36}\mbox{Ca})=2.56$~MeV,
$S_n(^{36}\mbox{S})=9.89$~MeV.}.
The dominant component of SMEC g.s. wave function is: 
\begin{eqnarray}
&&[(d_{5/2})^6(s_{1/2})^2]+[[(d_{5/2})^6(s_{1/2})^1]_{1/2^+}\otimes(s)^1_{(c)}]+  \\ 
&&[[(d_{5/2})^5(s_{1/2})^2]_{5/2^+}]\otimes(d)^1_{(c)}]+\cdots \nonumber
\end{eqnarray}
, where index $c$ denotes continuum state. Important contributions to $E^{\rm (corr)}_{\rm 2_1^+}$ 
are spread over six states in daughter nuclei. Contributions from couplings to the two lowest states ($1/2^+_{1}$ and $3/2^+_{1}$) favor $^{36}\mbox{Ca}$ state, whereas all contributions from couplings to higher excited states ($7/2^+_{1}$, $3/2^+_{2}$, $5/2^+_{2}$ and $9/2^+_{1}$) favor $^{36}\mbox{S}$ state.  As a result, the $2_1^+$ state in $^{36}\mbox{Ca}$ is shifted down with respect to $^{36}\mbox{S}$. The dominant component of the SMEC $2_1^+$ wave function is: 
\begin{eqnarray}
&&[(d_{5/2})^6(s_{1/2})^1(d_{3/2})^1]+
[[(d_{5/2})^6(s_{1/2})^1]_{1/2^+}\otimes(d)^1_{(c)}]+\nonumber \\ 
&&[[(d_{5/2})^6(d_{3/2})^1]_{3/2^+}\otimes(s)^1_{(c)}]+\\ \nonumber
&&[[(d_{5/2})^5(s_{1/2})^1(d_{3/2})^1]_{9/2^+}\otimes(d)^1_{(c)}]+
[[(d_{5/2})^5(s_{1/2})^1(d_{3/2})^1]_{7/2^+}\otimes(d)^1_{(c)}]+\\ \nonumber
&&[[(d_{5/2})^5(s_{1/2})^2]_{5/2^+}\otimes(s)^1_{(c)}]+
[[(d_{5/2})^5(s_{1/2})^1(d_{3/2})^1]_{3/2^+}\otimes(s)^1_{(c)}]+\cdots
\nonumber 
\label{dwaplus}
\end{eqnarray}

If one looks to the $2_1^+ - 0_1^+$ energy difference in $^{36}\mbox{Ca}$ and $^{36}\mbox{S}$, one may notice that a contribution to the energy of $0_1^+$ from a coupling to $1/2^+_{1}$ in $A-1$ nuclei is almost exactly compensated by contributions to the energy of $2_1^+$ from couplings to $1/2^+_{1}$ and  $3/2^+_{1}$ states. The asymmetry in the position of $2_1^+$ states in  
$^{36}\mbox{Ca}$ and $^{36}\mbox{S}$ comes from the balance between the contribution of the $5/2^+_{1}$ in the g.s. and contributions of higher lying states  $7/2^+_1$, $3/2^+_2$, $5/2^+_2$ and $9/2^+_1$ in the $2_1^+$ state. These two sets of couplings act differently in the g.s. and in the first excited state. It is interesting to notice that the effect is produced by continuum couplings which practically do not involve $s$-wave neutron/proton components. These components, which are present both in the $0_1^+$ state and in the $2_1^+$ state (cf Eq. (4)), mutually cancel out and do not contribute to the Thomas-Ehrmann shift in  $^{36}\mbox{Ca}$-$^{36}\mbox{S}$ mirror systems.

The proximity of $fp$ shell leads to important excitations from $sd$ to $fp$ across the $N=20$, $Z=20$ shell closure.  As  a consequence, many new channels $[J_i^{\pi;A-1};(l,j)]^{J_{\rm g.s.}^{\pi;A}}$ become involved in the continuum coupling, leading to an enhanced spreading of contributions to the continuum-coupling energy correction. In general, an increased spreading of couplings leads to an increased fractionation of  continuum-coupling contributions and the reduction of $E^{\rm (corr)}$ due to an enhanced interference of large number of channels. 

\begin{table}[h]
\caption{Fraction of $0\hbar\omega$ configurations in $2\hbar\omega$
  SM calculations for $^{36}\mbox{Ca}$/$^{36}\mbox{S}$ ($A=36$) and corresponding daughter nuclei 
  ($A=35$).  Results correspond to $0_1^+$, $2_1^+$ states of the parent nucleus ($A=36$) and to averages over 7 (resp. 8) most important states in $T_z=5/2$ (resp. $T_z=3/2$) daughter nuclei, respectively. }
 \label{table2}
\begin{center}
\begin{tabular}{|c|c|c|c|c|c|}
\hline
 Hamilt. & space & $0_1^+(A=36)$ & $2_1^+(A=36)$ & 
 ${\langle T_z=5/2\rangle}_{\rm av}$& ${\langle T_z=3/2\rangle}_{\rm av}$ \\
\hline
 IOKIN & sdfp & 0.769 & 0.768 & 0.772 & 0.782 \\
 WBT   & sdfp & 0.787 & 0.789 & 0.795 & 0.796 \\
 WBT  & psdfp & 0.687 & 0.684 & 0.683 & 0.685 \\
 \hline
\end{tabular}
\end{center}
\end{table}  
\begin{figure}[h]
\psfig{file=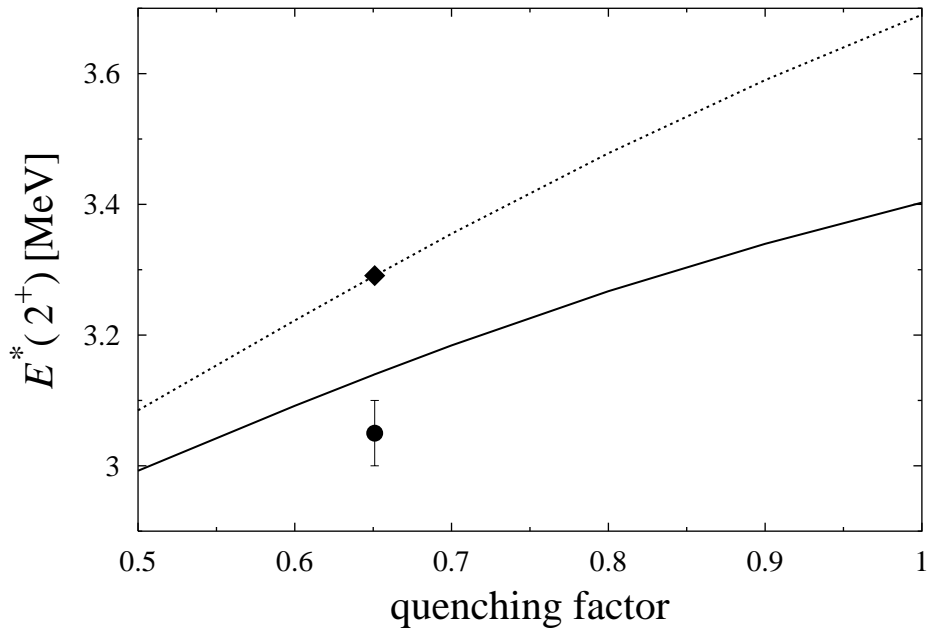,width=9cm,angle=-90}
\caption{SMEC excitation energies of $2_1^+$ states in $^{36}\mbox{Ca}$ and
  $^{36}\mbox{S}$ are plotted as a function of the quenching factor.
  Experimental excitation energies of $2_1^+$ state in $^{36}\mbox{S}$ 
  and $^{36}\mbox{Ca}$ are depicted as diamond and circle, respectively. }
\label{fig2}
\end{figure}
Full SMEC calculations in $sdfp$ shells for $^{36}\mbox{Ca}$ and $^{36}\mbox{S}$ are beyond actual possibilities so we proceed by introducing the quenching factor  $Q_{\rm f}$ in the continuum-coupling energy correction \cite{bnop2}. SM calculations using WBT \cite{warburton} and IOKIN \cite{nummela} effective Hamiltonians, which include $fp$ shell and allow for $2\hbar\omega$ excitations, revealed that an appropriate quenching factor should be used to account for admixture of intruder configurations (cf Fig. \ref{table2}). Almost identical admixtures of intruder configurations have been found both in parent nuclei and in all important states of $A-1$ nuclei.  

The value of the quenching factor depends on different extensions of $sd$-shell effective interaction  into a larger model space (cf Table 2).  It turns out that the excitation energies of $2_1^+$ states in $^{36}\mbox{Ca}$ and  $^{36}\mbox{S}$ are sensitive to the value of the quenching factor (see Fig.~\ref{fig2}). Both absolute excitation energies and their difference diminish as the quenching factor decreases. For $Q_{\rm f}=0.651$, the SMEC excitation energy of $2_1^+$ state reproduces exactly the experimental value in well-bound $^{36}\mbox{S}$. This value for the quenching factor  is close to the value found for WBT Hamiltonian (cf Table \ref{table2}) in $psdfp$ shells.  

For IOKIN and WBT Hamiltonians in $sdfp$ shells, one obtains higher values for $Q_{\rm f}$. For those values, SMEC reproduces well an experimental difference of $2_1^+$ energies in  $^{36}\mbox{Ca}$ and $^{36}\mbox{S}$ nuclei, but the absolute values of $2_1^+$ energies are too high. The latter problem could be easily resolved by mirror-symmetry conserving correction of the  $V^{T=1}_{0d_{5/2};1s_{1/2}}$ monopole.

\section{Conclusions}
In this paper, we attempted to address two problems pertaining to the physics with exotic, weakly-bound nuclei: (i) What are the generic features of the continuum coupling in binding systematics of neutron-rich nuclei? (ii) Can one explain mirror-symmetry breaking effects in spectra by the asymmetry of threshold energies  in mirror  nuclei?

The answers to these questions are not simple and require further investigations. Studies in long chains of oxygen and fluorine isotopes revealed that the $np$ residual coupling to the scattering continuum becomes strongly reduced with respect to the $nn$ coupling in the vicinity of the neutron drip line. The $np$ coupling is essential for an understanding of the anti-OES effect which is seen in odd-$Z$ fluorine chain and leads to an apparent reduction of the gap parameter for neutrons. Hence, in weakly-bound nuclei close to the drip lines, the OES has three components: the first one originates from nucleonic pairing, the second one is the deformed mean-field effect \cite{satula}, and the third one originates from the $np$ coupling via scattering continuum. One should stress that the latter component  is of a completely different nature than the singular behavior of binding energies originating from the $np$ correlations near $N=Z$ line \cite{nz}.

Asymmetry in the spectra of mirror systems is another playground for the continuum shell model. In the mirror couple $^{36}\mbox{Ca}$-$^{36}\mbox{S}$, a relative shift of $2_1^+$ excitation energies can be to a large extent directly related to the effect of the continuum coupling. However, in contrast to neutron-rich nuclei, the low one-proton separation energy in $^{36}\mbox{Ca}$ has little influence on the difference of excitation energies of $2_1^+$ states in  $^{36}\mbox{Ca}$ and $^{36}\mbox{S}$. This is due to the Coulomb barrier which suppresses 
continuum-coupling effects in weakly-bound systems close to the proton drip line. In that respect, weakly-bound/unbound systems at the proton drip line are radically different from those at the neutron drip line. To understand dissimilarity of nuclear systems at the proton and neutron drip lines is a challenge for the nuclear structure theory.

\begin{acknowledgments}
One of us (M.P.) wish to thank F. Azaiez for stimulating discussions.
\end{acknowledgments}

\end{document}